\documentclass[10pt, journal]{IEEEtran}
\IEEEoverridecommandlockouts
\usepackage{amsmath,graphicx,amssymb,array,multirow,cite,float,setspace,xcolor}
\usepackage[hyphens]{url}
\newcommand{\rt}[1]{\textcolor{black}{#1}}

\title{Forward Collision Vehicular Radar with \\IEEE 802.11: Feasibility Demonstration through Measurements}
\author{Robert C. Daniels, Enoch R. Yeh \emph{Student Member, IEEE}, and Robert W. Heath, Jr., \emph{Fellow, IEEE}\\
\thanks{This research was partially supported by the U.S. Department of Transportation through the Data-Supported Transportation Operations and Planning (D-STOP) Tier 1 University Transportation Center.  This work was also sponsoblack by the Texas Department of Transportation under Project 0-6877 entitled âCommunications and Radar-Supported Transportation Operations and Planning (CAR-STOP).  This material is based upon work supported in part by the National Science Foundation under Grant No. NSF-1549663.  This work was additionally supported by Toyota. \newline \indent E. R. Yeh and Dr. R. W. Heath Jr. are with the Wireless Networking and Communications Group in the Department of Electrical and Computer Engineering of the University of Texas at Austin.  \newline \indent Dr. R. C. Daniels is with PHAZR, Inc.}}

\begin{document}
\maketitle


\begin{abstract}

Increasing safety and automation in transportation systems has led to the proliferation of radar and \rt{IEEE 802.11p-based} dedicated short range communication (DSRC) in vehicles.  Current implementations of vehicular radar devices, however, are expensive, use a substantial amount of bandwidth, and are susceptible to multiple security risks.  In this paper, we use the IEEE 802.11 orthogonal frequency division multiplexing (OFDM) communications waveform, \rt{as found in IEEE 802.11a/g/p}, to perform radar functions.  In this paper, we present an approach that determines the mean-normalized channel energy from frequency domain channel estimates and models it as a direct sinusoidal function of target range, enabling closest-target range estimation.  In addition, we propose an alternative to vehicular forward collision detection by extending IEEE 802.11 dedicated short-range communications (DSRC) and WiFi technology to radar, extending the foundation of joint communications and radar frameworks.  Furthermore, we perform an experimental demonstration near DSRC spectrum using IEEE 802.11 standard compliant sotware defined radios with \textcolor{black}{potentially} minimal modification through algorithm processing on frequency-domain channel estimates.  \textcolor{black}{The results of this paper show that our solution delivers sufficient accuracy and reliability for vehicular RADAR if we use the largest bandwidth available to IEEE 802.11p ($20~\mathrm{MHz}$).} This indicates significant potential for industrial devices with joint vehicular communications and radar capabilities.

\end{abstract}

\section{Introduction}
\label{sec:intro}
Radar is a key technology for frontal collision avoidance.  \textcolor{black}{Vehicular radars have several advantages over competing sensor technologies.  For example, although considerable progress has been made on improving light detection and ranging (LIDAR) sensors, they are still expensive relative to vehicular radar and exhibit sensitivity issues in low-visibility conditions \cite{Naughton2017,SmithPeter2000}, ultrasonic sensors have limited range and suffer from degraded performance in several common environments \cite{Carullo2001,Klotz2000}, and dedicated short range communication (DSRC) networks (which can be used to exchange GPS-based location and velocity information) only work when the colliding vehicle has a radio \cite{Kenney2011}.}

Despite their importance in vehicle safety, existing forward vehicular radar products have several limitations.  First, current vehicular radars are implemented at high carrier frequencies and require large bandwidth allocations (i.e., hundblacks of MHz), forcing them to exploit the millimeter wave (mmWave) spectrum.  mmWave radars feature complex antenna array structures and high power analog hardware, leading to increased cost relative to IEEE 802.11 transceivers \cite{Hasch2012}.  This affects the original equipment manufacturers (OEMs) of economy vehicles that deploy vehicular radars as well as consumers needing to replace damaged parts, since radars are mounted on the front bumper of vehicles and are susceptible to collision damage.  Second, mmWave radar antennas cannot be decoupled from the analog and RF circuits due to cable losses at mmWave frequencies \cite{Rappaport2014}.  Consequently, all of the circuitry of mmWave radars is contained within a single package which leads to either degrading aerodynamics and visual appeal or requiring extra effort to conceal when placed on vehicles \cite{Cerretelli2007}.  \textcolor{black}{Third, existing vehicular radars are known to be susceptible to certain security risks such as spoofing, which enables malicious attackers to impose false readings and cause accidents \cite{Hubaux2004,Hasch2012,Chauhan2014}}.  Traditional vehicular mmWave radars leverage a specific signal structure with strong autocorrelation properties but exhibits no inherent authentication \cite{Yeh2016}.  Without additional authentication from a supplemental state-of-the-art communication network, these radars remain vulnerable to spoofing attacks.  Fourth, because there is no uniformly-adopted standard for vehicular radar waveforms, electromagnetic interference may eventually limit the scale of deployment \cite{Brooker2007}.

The proliferation of dual radio devices, however, supports the use of IEEE 802.11 communication networks for vehicular safety \cite{Sturm2011}.  In addition, the Federal Communications Commissions (FCC) has allocated spectrum at the $5.9~\mathrm{GHz}$ band specifically for DSRC as part of IEEE 802.11p.  Vehicles equipped with DSRC for active safety applications have already been deployed in Japan and are ready for deployment in the United States \cite{TTT2015,USDOT2015}.

Successful integration of ranging capabilities into existing IEEE 802.11 wireless communication devices provides substantial opportunities for automotive OEMs and its customers.  Attaching sufficient ranging capabilities to such devices, however, is challenging due to bandwidth limitations.  Current frequency-modulation continuous wave (FMCW) vehicular radars with digital receiver processing require $150~\mathrm{MHz}$ of bandwidth for $1~\mathrm{m}$ range resolution \cite{Kihei2015,Rohling2001}. \textcolor{black}{In contrast, IEEE 802.11p waveforms only operate on $5~\mathrm{MHz}$, $10~\mathrm{MHz}$, or $20~\mathrm{MHz}$ spectrum allocations and IEEE 802.11a/g/n waveforms often only occupy $20~\mathrm{MHz}$ or $40~\mathrm{MHz}$ spectrum allocations (IEEE 802.11ac has optional $80~\mathrm{MHz}$ and $160~\mathrm{MHz}$ modes \cite{Van2011}).}

There has been considerable research on passive bistatic and multistatic radar with IEEE 802.11 packets demonstrating refined single and multiple-target localization in indoor and outdoor environments \cite{Colone2012,Falcone2012,Chetty2012,Maechler2012,Ivashko2014,Rzewuski2013}.  Passive radar, though, is not suitable for automotive safety applications since it requires a high density of infrastructure.  Active high-resolution radar has recently been demonstrated for through-the-wall human detection (through measurements) and vehicular radar (through simulation) using IEEE 802.11 OFDM waveforms \cite{Adib2013,Letzepis2011}.  Both of these methods rely on a complete blackesign of the digital baseband receiver \cite{Letzepis2011}.  Unfortunately, this will delay product deployment and partially erode the cost benefit afforded to IEEE 802.11 based radars, making widescale deployment of active high-resolution radars impractical.  


\textcolor{black}{Also known as the Boomerang Transmission System, \cite{Mizui1993} laid the foundations for V2V ranging by using spread-spectrum.  Since then, there have been several monostatic radar approaches that leverage OFDM signal structure in IEEE 802.11 packets.  Some approaches, such as \cite{Sturm2010,Sit2011radar,Braun2009}, fail to achieve the accuracy for long-range radar applications as specified in \cite{Hasch2012}}.  Other OFDM radar approaches require large amounts of bandwidth unavailable to DSRC \cite{Garmatyuk2009,Crouse2012,Sit2014,Kashin2013,Sen2011}.  While \cite{Reichardt2012} claims to have achieved accuracy of less than $2$ m for IEEE 802.11p OFDM vehicular radar using traditional radar techniques, they require an additional $60 ~\mathrm{MHz}$ spectrum from an adjacent ISM band, which may not be available for use in the deployment stage.  \textcolor{black}{A suggested framework for IEEE 802.11p radar in \cite{Kihei2015} models the 802.11p signal as a multi-frequency continuous waveform (MFCW) to estimate range and velocity of a vehicular target; however, this approach lacks experimental validation.  CohdaWireless claims that they have achieved a non-intrusive 360-degree vehicular radar using frequency-domain channel estimates from an NXP Roadlink chipset using a correlator approach \cite{Alexander2015}.  They claim to have demonstrated GPSless positioning at $5.9 \mathrm{~GHz}$, using V2X hardware deployment and over-the-air messaging to relay GPS coordinates \cite{CohdaWireless2016}.  This demonstration, however, does not explicitly address ranging accuracy, which is crucial for collision avoidance technology.  As a result, we believe that our paper provides the first feasibility study of this technology for collision avoidance systems.}


In this paper, we design a simple algorithm for computing high-resolution ranging and target detection metrics from frequency domain channel estimates obtained from an IEEE 802.11 OFDM receiver without modification to the digital baseband receiver.  We assume that the return channel can be well-modeled by the two-path model, that is, the receive signal consists of a direct path and a single reflected path.  Under this assumption, the mean-normalized channel energy computed from the frequency domain channel estimates can be well-approximated by a direct sinusoidal function of target range.  This enables us to directlranging.texy estimate target range through parameter optimization of sinusoidal functions.  We apply a three-dimensional brute-force optimization algorithm to estimate the time delay associated with the reflected path, which directly corresponds to the range of the closest target.

In simulation, we show that meter-level accuracy can be achieved using IEEE 802.11 packets with a $20~\mathrm{MHz}$ channel bandwidth.  We validate our simulation results through measurements using a prototype IEEE 802.11a transceiver over-the-air with vehicle targets.  This prototype was implemented with an off-the-shelf software radio transceiver and two directional antennas.  Using standard IEEE 802.11a transmission, our prototype achieves meter-level accuracy with meter-level resolution for a single vehicle target up to $50$ m away from the \rt{radar transceiver}.

The immediate consequences of our study are threefold.  First, since IEEE 802.11 networks are already being transferblack into automotive environments for dedicated short-range communications (DSRC) through IEEE 802.11p \cite{Jiang2008}, we have essentially enabled immediate market access to dual purpose radar and communication (RadCom) devices \cite{Sit2011}.  This benefit is strengthened by the observation that mmWave communication systems, the previous communications standard candidate proposed for RadCom \cite{Kumari2015}, are only starting to penetrate the marketplace \cite{Rappaport2014}.  Second, our study proves that cost-effective vehicular radar solutions in a convenient form factor are possible.  Finally, our results show that a new sophisticated radar waveform and medium access control (MAC) layer standard is not requiblack to provide existing vehicular radar networks with coexistence, security, and power management features, \rt{since these are achieved naturally by IEEE 802.11} \cite{Jiang2008}.

The remainder of this paper is organized as follows. Section \ref{sec:system} describes the system model for the IEEE 802.11 channel and summarizes key assumptions.  Section \ref{sec:ranging} discusses the algorithmic approach used to determine range estimation based on receive channel energy.  Section \ref{sec:sims} presents simulation results and discusses the impact of these results on the overall study.  Section \ref{sec:platform} describes the measurement platform of the study. Section \ref{sec:scenarios} discusses the measurement method and environment and presents the results of the study. Section \ref{sec:conclusion} discusses the implications of the results, concludes the paper, and provides suggestions for future work.

\emph{Notation}: The real coordinate space with $N$ dimensions is denoted by $\mathbb{R}^N$. The natural coordinate space is denoted by $\mathbb{N}$.

%
\section{System Model}
\label{sec:system}
In this study of IEEE 802.11-based vehicular ranging feasibility, we assume that a vehicle is equipped with an IEEE 802.11 transceiver.  The transmit and receive antennas exhibit sufficient separation such that a direct signal path from transmit to receive antenna exists.  Further, we assume that the receive RF chain is able to operate simultaneously with the transmit chain.  This configuration can be found in commercial IEEE 802.11 devices, for example, to enable channel reciprocity through calibration.  As such, our algorithm will operate under the assumption of a pseudo-monostatic active radar configuration.   In addition, we assume that signals can be transmitted at a rate adequate for performing range detection in a constantly fluctuating vehicular environment.

Consider the continuous-time complex-baseband link model with time index $t$, transmitted signal $x(t)$, received signal $y(t)$, wireless channel impulse response $h(t)$ with $\tau$ excess delay, and additive noise signal $v(t)$ such that $y(t) = h(t) \ast x(t) + v(t)$ where $\ast$ is the convolution operator.  For IEEE 802.11, it is generally assumed that the channel impulse response will remain fixed over the duration of a single packet, but may change between packets due to mobility.

\begin{figure}
  \begin{center}
    \includegraphics[width=2.5in]{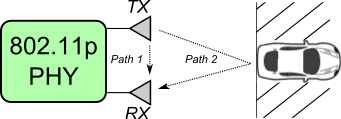}
    \caption{Illustration of system configuration.  The IEEE 802.11p physical layer (PHY) is connected to separate transmit (TX) and receive (RX) RF chains. \textcolor{black}{Path 1 is the direct path which may result from antenna sidelobes or any leakage between the transmit and receive chains.} Path 2 is the reflected path from the ranging target.}
    \label{fig:two_path_model}
  \end{center}
\end{figure}

Now, we simplify our general link model to a two-path channel model to address the vehicular ranging application: forward collision detection with a primary target (the vehicle directly ahead).  \textcolor{black}{Our application of this model from an experimental perspective is not unique. It is common to create a reference response to evaluate the cross-correlation function of the reflection responses\cite{Colone2012}. In our case, we simplify the reflection responses to a single reflection to maximize tractability.} We adopt this model under the assumption that we are able to effectively create a direct path between transmit and receive antennas for monostatic radar.  This special case is illustrated in Figure \ref{fig:two_path_model}. 
   
Consider the pseudo-complex baseband channel equivalent \cite{Rappaport2014}
\begin{equation}
h_{p} (t) = \alpha \delta(t) + \beta e^{-j (2 \pi f_c \tau - \phi_0)} \delta(t -\tau).
\end{equation}
where $\delta(\cdot)$ is the Dirac delta function, $f_c$ is the carrier frequency, $\phi_0$ is the constant phase shift of the reflected signal with respect to the direct signal, $\tau$ becomes the time delay associated with target reflection, $\alpha$ is the path loss associated with the direct path (Path 1), and $\beta$ is the path loss associated with the reflected target (Path 2).  Note that since we use the relative phase shift between the direct and reflected path $\phi_0$, the phase shift of the reflected direct path is omitted.  The direct path behaves as a reference point for the target reflection path, enabling accurate estimation of the delay between the direct path arrival and the target reflection arrival.  The path loss of the direct path, $\alpha$, is a function of the transmit power ($P$), the path loss coefficient of the direct path between the transmit and receive antenna ($L_1$), the signal power feed-through coefficient between the transmit and receive paths in the analog/RF circuit ($F$), and the gain based on radiation pattern in the unintended direction of the transmit and receive antennas ($G^{(\text{TX})}_1$ and $G^{(\text{RX})}_1$, respectively) such that
\begin{equation}
        \alpha = \sqrt{ P F } + \sqrt{ P G^{(\text{TX})}_1 G^{(\text{RX})}_1 L_1 }.
\end{equation}
Note that IEEE 802.11a/g/p transceivers generally experience signal power feed-through within the analog/RF circuit.  Our model assumes that the direct path is a combination of the effects from internal feed-through and external line-of-sight propagation.

Similarly, the path loss of the reflected path, $\beta$, is a function of the transmit power ($P$), the single-direction path loss coefficient between the transmit antenna and the reflecting object ($L_{2,1}$), the single-direction path loss coefficient between the reflecting object and the receive antenna ($L_{2,2}$), the reflection power loss coefficient ($R$), and the gain based on the radiation pattern of the transmit and receive antennas ($G^{(\text{TX})}_2$ and $G^{(\text{RX})}_2$, respectively) such that
\begin{equation}
        \beta = \sqrt{ P G^{(\text{TX})}_2 G^{(\text{RX})}_2 L_{2,1} L_{2,2} R }.
\end{equation}
The Friis transmission equation determines the path loss coefficients $L_1$, $L_{2,1}$, and $L_{2,2}$. The radar cross section (RCS) equation determines the reflection power loss coefficient $R$ as a function of wavelength ($\lambda$) and a RCS parameter ($\sigma$):
\begin{equation}
R = \frac{ 4 \pi \sigma }{ \lambda^2 }.
\end{equation}

For bandwidth $B$, the continuous-time channel impulse response can then be represented as
\begin{equation}
h(t) = \alpha B ~\mathrm{sinc} (Bt) + \beta e^{-j 2\pi f_c \tau} B ~\mathrm{sinc} (B(t - \tau))
\end{equation}
where $\mathrm{sinc}(x) = \sin(\pi x)/\pi x$.
Assume that the receive signal is perfectly bandlimited.  For symbol period $T = 1/B$, the discrete-time channel impulse response is
\begin{align}
h[n] & = T h(nT) \\
& = \alpha ~\mathrm{sinc} (Btn) + \beta e^{-j 2\pi f_c \tau - \phi_0} \mathrm{sinc} (BTn - B\tau) \\
& = \alpha \delta [n] + \beta e^{-j 2\pi f_c \tau -\phi_0} \mathrm{sinc} (n - B\tau). \label{eq:impulse}
\end{align}
For frequency $f$, the discrete-time Fourier transform (DTFT) of (\ref{eq:impulse}) is 
\begin{equation}
H(e^{j 2\pi f}) = \alpha + \beta e^{ -j ( 2 \pi f_c \tau - \phi_0) }e^{ -j ( 2 \pi f \tau - \phi_1) }
\end{equation}
where $\phi_1$ is the frequency varying phase shift of the reflected signal with respect to the direct signal and $f \in [-\frac{1}{2}, \frac{1}{2}]$.

IEEE 802.11a/g/p standards use OFDM; thus, frequency domain channel estimates enabled by the discrete Fourier transform (DFT) are typically provided by OFDM transceivers. Assuming perfect synchronization, perfectly band-limited signals, and perfect estimation algorithms, if $\tau$ is smaller than the duration of the OFDM cyclic prefix, for subcarrier bandwidth $\Delta$ and $N$ nonzero subcarriers $m\in\{-N/2, -N/2+1,\dots,N/2\},m\ne 0$, the baseband OFDM frequency domain channel estimate of the two-path channel, obtained by sampling the DTFT at each subcarrier, is \cite{Marchetti2009}
\begin{equation}
\hat{H}[m] = \alpha + \beta e^{ -j \left( 2 \pi m \Delta \tau - \theta \right) }.
\end{equation}
where $\theta = \phi_0 + \phi_1$ is the overall phase shift of the reflected signal with respect to the direct signal.  For the sake of notation, we will omit the carrier frequency term.  Note that practical channel estimates will include noise and various filter contributions.  

To determine the mean-normalized channel energy, we define the magnitude of the channel coefficients, which we call the energy of the channel estimates, as
\begin{eqnarray}
E_{\hat{H}}[m] & = & |\hat{H}(m \Delta)|^2  \\
& = & \alpha^2 + \beta^2 + 2\alpha\beta\cos\left( 2 \pi m \Delta \tau - \theta \right).
\end{eqnarray}
Finally, we define the empirical mean over all subcarriers as
\begin{equation}
\tilde{E}_{\hat{H}} = \frac{1}{N} \sum \limits_n E_{\hat{H}}[n]
\end{equation}
and compute the mean-normalized channel energy
\begin{equation}
\overline{E}_{\hat{H}}[m] = \frac{E_{\hat{H}}[m]}{\tilde{E}_{\hat{H}}}.
\end{equation}
For the two path channel,
\begin{align}
\overline{E}_{\hat{H}}[m] - 1 ~ & = ~\frac{\alpha^2 + \beta^2 + 2\alpha\beta\cos\left( 2 \pi m \Delta \tau - \theta \right)}{\frac{1}{N} \sum \limits_n  \alpha^2 + \beta^2 + 2\alpha\beta\cos\left( 2 \pi n \Delta \tau - \theta \right)} - 1 \label{eq:comp}\\
& \stackrel{(a)}{\approx} ~\frac{2\beta}{\alpha} \cos\left(2\pi m \Delta \tau - \theta \right) \label{eq:metric}
\end{align}
where (a) follows from $\alpha \gg 2\alpha\beta \gg \beta$.  This enables us to directly estimate target range by determining the delay $\tau$ from the adjusted channel estimate energy.

Recall that we have omitted the phase shift of the direct path in our channel model.  For phase shifts $\theta_d$ and $\theta_r$ of the direct and reflected paths respectively, consider briefly blackefining the channel frequency response as
\begin{align}
\tilde{H}[m] & = \alpha e^{j \theta_d} + \beta e^{ -j \left( 2 \pi m \Delta \tau - \theta_r \right) } \\
& = e^{j \theta_d}(\alpha + \beta e^{ -j \left( 2 \pi m \Delta \tau - \theta \right) })
\end{align}
where the relative phase shift of the reflected path has the relation $\theta = \theta_r - \theta_d$.  Now the mean-normalized channel energy would be
\begin{align}
E_{\tilde{H}}[m] & = |\tilde{H}(m \Delta)|^2  \\
& = \big|e^{j \theta_d}\big|^2 (\alpha^2 + \beta^2 + 2\alpha\beta\cos\left( 2 \pi m \Delta \tau - \theta \right)) \\
& = \alpha^2 + \beta^2 + 2\alpha\beta\cos\left( 2 \pi m \Delta \tau - \theta \right)
\end{align}
which is the same as before.  Thus, to simplify notation, we have omitted the phase shift of the direct path.

Mobile OFDM systems exhibit Doppler effects, which generates inter-carrier interference between OFDM subcarriers and degrades performance \cite{Wang2006}.  Estimates of the Doppler spread can be used to mitigate the negative effects of a highly mobile wireless channel \cite{Yucek2005}.  In addition, Doppler is a critical parameter for estimating target velocity in radar \cite{Stove1992}.  Standard radar methods for Doppler frequency estimation require long transmissions since waveforms are coherently processed for many consecutive milliseconds (ms) \cite{Rohling2001}.  IEEE 802.11a/g/p packets, however, are rarely more than $1$ ms in duration, and often much shorter, making it difficult to directly estimate Doppler from a single packet.  Nevertheless, target velocity can be estimated between packets through differential computations on ranging estimates from packets. Doppler frequency is not expected to significantly distort channel estimates \rt{since the channel is estimated using only the pair of long training OFDM symbols which are part of the physical layer converge protocol preamble.} The duration of one long training OFDM symbol of the IEEE 802.11p waveform is $6.4$ microseconds ($\mu$s) and the Doppler frequency in vehicular environments, assuming a maximum differential velocity of $160~ \textrm{km/h}$ (\textcolor{black}{two cars approaching each other at $80~\textrm{km/h}$}), does not exceed $875$ Hz \cite{Cheng2008}.  Hence, channel estimation is based on training sequences in the long training symbols whose worst-case Doppler is less than 6 degrees per OFDM symbol per target. \rt{ It may be possible to improve performance by estimating the channel using all of the transmitted OFDM symbols, by exploiting the fact their contents are known since the transmitter and receive are co-located.  We defer this to future work, as it departs from standard communication receiver processing that only uses the long OFDM training symbols for channel estimation. As a result, for the scope of this feasibility study, we do not explicitly consider Doppler effects in our model and reserve the topic of velocity detection for future work. }
  \textcolor{black}{Note that all of the simplifying assumptions in this section, i.e., perfect synchronization, bandlimited signals, etc., only apply to the system model in this section and the algorithm in the sequel.  The implemented algorithm and experimental results in later sections suffeblack from all of the practical imperfections that would be found in a real IEEE 802.11 radio.  Future work may systematically model these impairments to further reinforce our feasibility study.}

\section{Algorithm for Ranging and Detection}
\label{sec:ranging}
In this section, we describe an algorithm for range estimation that applies a brute-force minimization of the least-squablack error between sinusoidal functions, which is a parametric estimation.  In various radar implementations, spectrum estimation is used to determine ranging based on a channel impulse response \cite{Welsh1967,Gabriel1986}.  This strategy is effective in many existing antenna system applications, including MIMO radar systems \cite{Byrnes2000,Xu2008}.  We chose brute-force optimization over alternative procedures for two primary reasons: (1) We are interested in testing the feasibility of vehicular ranging with our proposed approach.  A parameterized brute-force approach is a good baseline for system performance and ensures that any performance limitations discoveblack are not due to the optimization algorithm.  (2) Vehicular radar estimates are performed over durations large enough such that algorithm complexity is not likely to be a limiting factor (e.g., $50~\mathrm{ms}$ update interval on Delphi Electronic Scanning radars \cite{DelphiESR}).  Future work should examine the effect of alternative optimization procedures such as spectrum estimation to blackuce algorithm complexity.


\subsection{Model Error Minimization}
\label{sec:ranging1}
Based on (\ref{eq:metric}) in Section \ref{sec:system}, we propose a brute-force optimization algorithm that matches a sinusoid to the received mean-normalized channel energy.  Consider mean offset ($A \in \mathbb{R}$), cosine magnitude candidate ($B \in \mathbb{R}$), initial phase candidate ($C \in [0, 2\pi)$), and phase increment ($D \in \mathbb{R}$).  Using (\ref{eq:metric}), we can model the estimated mean-normalized channel energy $\hat{x}[m]$ as
\begin{equation}
\hat{x}[m] = A + B \cos(C + Dm).
\end{equation}
Given the true received mean-normalized channel energy $x[m]$, the error function
\begin{eqnarray}
\epsilon(x[m], \hat{x}[m]) &=& |\hat{x}[m] - x[m]|^2 \label{eq:error_func} \\
& = &  \bigg|\hat{x}[m] - \frac{2\beta}{\alpha} \cos\left(2\pi m \Delta \tau - \theta \right)\bigg|^2
\end{eqnarray}
is minimized when the parameters of $\hat{x}[m]$ correspond to those of $x[m]$, i.e., $A = 0$, $B = 2\beta/\alpha$, $C = -\theta$, and 
\begin{equation}
D = \frac{4 \pi f_s \rho}{cN}  = 2 \pi \Delta \tau
\label{eq:D_and_delay}
\end{equation}
assuming the received mean-normalized channel energy has a specific time delay and given PHY sample rate $f_s$,  $\rho$ is the target range candidate, $\tau$ is the time delay associated with the target reflection, and $c$ is the speed of light.  This algorithm performs minimization over three parameters: (1) the ratio of the path loss of the reflected path to the path loss of the direct path jointly represented by $A$ and $B$, (2) the phase offset $C$, and (3) the time delay associated with the target reflection $\tau$, which we will model as a function of target range candidate ($\rho \in [0,\infty)$). 

Brute-force three-dimensional minimization is implemented over a specified set of $A$, $C$, and $\rho$.  Cosine magnitude candidate $B$ is empirically determined from a sample of the received mean-normalized channel energy.  For each time delay $\rho$ in the set, the algorithm finds the path loss ratio $A$ and phase offset $C$ that returns the minimum least-squared error between the estimated mean-normalized channel energy $\hat{\mathbf{x}}$ (\rt{which collects the $x[m]$ corresponding to the $52$ subcarriers where the channel is estimated}) and the received mean-normalized channel energy $\mathbf{x}$ (\rt{which collects the $\hat{x}[m]$ corresponding to the $52$ subcarriers where the channel is estimated}), solving the minimization problem $\min \limits_{\hat{\mathbf{x}}} \left|\hat{\mathbf{x}} - \mathbf{x}\right|^2$.  \rt{In this way, our algorithm leverages the plurality of channel estimates on each OFDM subcarrier.} Once all phase increments $D$ corresponding to a set of ranges $\rho$ have been tested, the algorithm chooses the phase increment with the least residual error and returns the corresponding range $\rho$ as the range estimate.

The steps taken in this paper for range parameter determination from a single frequency domain channel estimate listed in order, are as follows.
\begin{enumerate}
        \item Define variables $A \in \mathbb{R}$, $B \in \mathbb{R}$, $C \in [0, 2\pi)$, $D \in \mathbb{R}$, $\rho \in [0, \infty)$, $x \in \mathbb{R}^N$, and $\hat{x} \in \mathbb{R}^N$ for iteration $k\in{\mathbb{N}}$ as specified above.  In addition, define the minimum residual error $\epsilon_{\mathrm{min}} \in \mathbb{R}$, and the target range candidate corresponding to minimum error $\rho_{\mathrm{min}}\in[0,\infty)$. 
        \item Define the predefined working set for $A$, $C$, and $\rho$ as $\mathcal{S}_A$, $\mathcal{S}_C$, and $\mathcal{S}_{\rho}= \{\rho_0, \rho_1, \ldots, \rho_M\}$, respectively, such that $A\in\mathcal{S}_A$, $C\in\mathcal{S}_{C}$, and $\rho\in\mathcal{S}_{\rho}$.  Let $0 \le k < M$ where $M$ is the number of elements in $\mathcal{S}_{\rho}$.  Define the residual error threshold constant $\epsilon_t$ such that if $|\hat{\mathbf{x}} - \mathbf{x}|^2 > \epsilon_t$ for all $\hat{\mathbf{x}}$, then no target is present within the channel.
        \item Set the received mean-normalized channel energy $\mathbf{x} = \overline{E}_{\hat{H}}-1$ and initialize $k = 0$ and $\epsilon_{\mathrm{min}} = \epsilon_t$.
        \item Use $\mathbf{x}$ to empirically determine $B$.  Ideally, $B$ would be set to the maximum magnitude of the mean-normalized channel energy in $\mathbf{x}$ to minimize error between $\mathbf{x}$ and $\hat{\mathbf{x}}$.  In practice, due to noise and model imperfections, our algorithm estimates the cosine magnitude by taking an order statistic of the magnitudes of the mean-normalized channel energy.
        \item Select $\rho_k$ in $\mathcal{S}_{\rho}$ and define the corresponding phase increment, $D = 4\pi f_s \rho_k/(cN)$.
        \item Iterate through every combination of $A \in \mathcal{S}_A$ and $C \in \mathcal{S}_C$.  Use $A$, $B$, $C$, and $D$ to define the metric value estimate $\hat{\mathbf{x}} = A + B \cos( C + Dm )$.  
        \item For each $\hat{\mathbf{x}}$, if $\left| \hat{\mathbf{x}} - \mathbf{x} \right|^2 < \epsilon_{\mathrm{min}}$ then redefine $\epsilon_{\mathrm{min}} = \left| \hat{\mathbf{x}} - \mathbf{x} \right|^2$ and $\rho_{\mathrm{min}} = \rho_k$.
        \item Increment $k = k + 1$ and repeat steps 5-8 until all elements in $\mathcal{S}_{\rho}$ are exhausted.
        \item If $\epsilon_{\mathrm{min}} < \epsilon_t$, a target has been determined to be present with target range of $\rho_{\mathrm{min}}$.
\end{enumerate}

\subsection{Verification}
\label{sec:ranging2}

To verify the estimation between $(\ref{eq:comp})$ and $(\ref{eq:metric})$, we quantify the error as a function of range.  The mean-normalized channel energy as defined in $(\ref{eq:metric})$ was simulated for a channel with a single target at ranges between 0.5 and 50 m at 0.5 m increments.  Assuming that all parameters except the delay $\tau$ are known, we determined the optimal delay $\hat{\tau}$ that minimizes the least-squares error between $(\ref{eq:comp})$ and the simulated mean-normalized channel energies.  We determined the optimal delay using the Nelder-Mead simplex method \cite{Nelder1965} with the actual delay as an initial guess and found the optimal range for each simulated mean-normalized channel energy using $(\ref{eq:D_and_delay})$.

\begin{figure}
 \centering
  \includegraphics[width=3.5in]{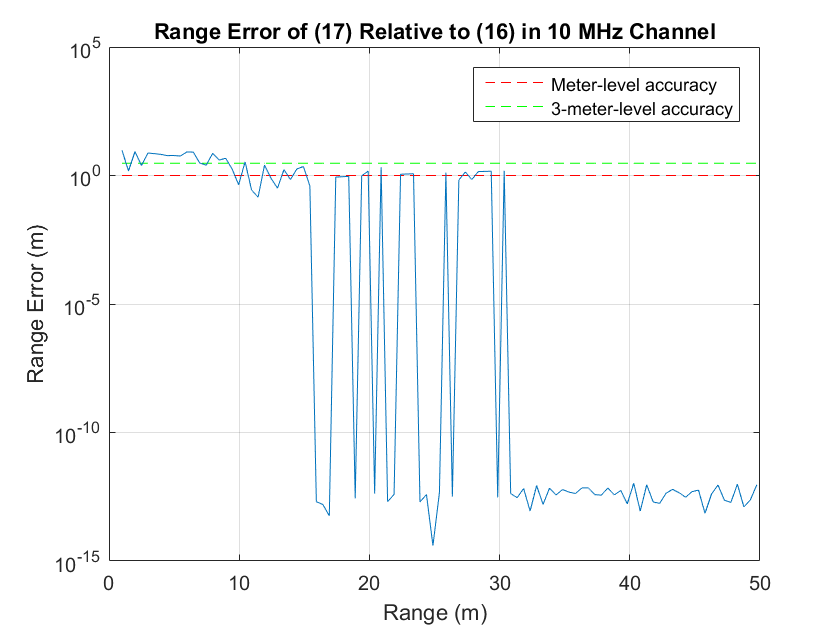}
  \caption{RMS range error between actual range of single target in a $10 ~\mathrm{MHz}$ channel and optimal range based on the simulated mean-normalized channel energy defined in $(\ref{eq:metric})$.  At $10 ~\mathrm{MHz}$, the estimation error does not exceed 3m when the target range is larger than 10m.}
  \label{fig:verify_10mhz}
\end{figure}

The plots for RMS range error between the actual range and optimal range found when simulating the mean-normalized channel energy using a $10 ~\mathrm{MHz}$ over 1000 Monte Carlo simulations are shown in Fig. \ref{fig:verify_10mhz}.  For a $10 ~\mathrm{MHz}$ channel, the error introduced by the estimation between $(\ref{eq:comp})$ and $(\ref{eq:metric})$ does not exceed 3 m when the range of the target is larger than 10 m.  In addition, as actual target range increases, the estimation generally becomes more accurate.  This implies that our model is more effective for long-range radar applications, which have a minimum detection distance of 10 m \cite{Hasch2012}.  Our estimation, however, fails to achieve the meter-level accuracy \rt{at those ranges} when using a $10 ~\mathrm{MHz}$ bandwidth.  As a result, we assess estimation performance using a $20 ~\mathrm{MHz}$ bandwidth.

For a $20 ~\mathrm{MHz}$ channel, the estimation error does not exceed 1 m when the range of the target is larger than 5 m.  Estimation accuracy improves as actual target range increases, similar to the $10 ~\mathrm{MHz}$ case.  The majority of the error in our range estimation algorithm described in Section \ref{sec:ranging1} should be from this estimation, indicating that we should be able to get very close to meter-level accuracy at single-target ranges greater than 5 m using a $20 ~\mathrm{MHz}$ channel.

\begin{figure}
  \centering
  \includegraphics[width=3.5in]{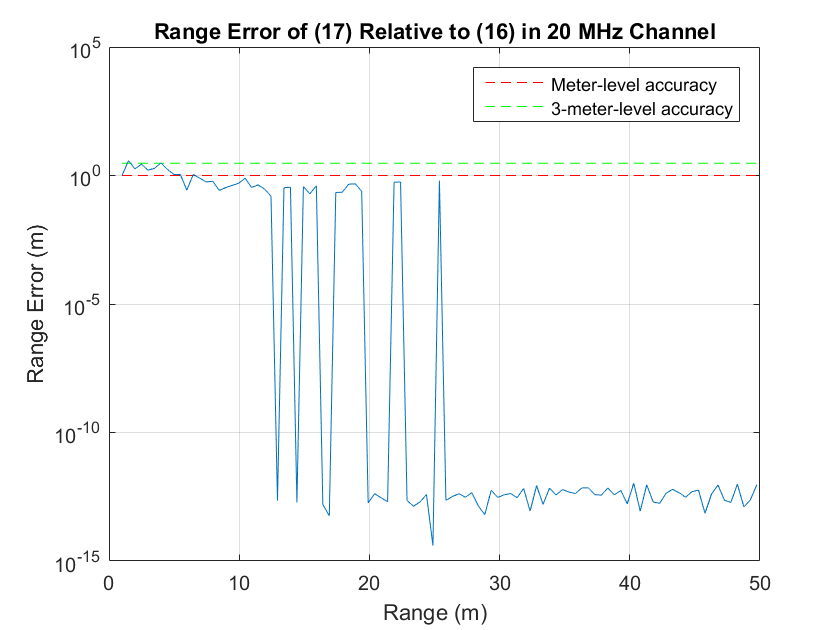}
  \caption{RMS range error between actual range of single target in a $20 ~\mathrm{MHz}$ channel and optimal range based on the simulated mean-normalized channel energy defined in $(\ref{eq:metric})$.  At $20 ~\mathrm{MHz}$, the estimation error does not exceed 1m when the target range is larger than 5m.}
  \label{fig:verify_20mhz}
\end{figure}

\subsection{Discussion}
\label{sec:ranging3}
Ideally, the brute force optimization does not need the offset parameters ($A$ and $C$).  Empirical results, however, have shown that a mean offset often occurs in practice.  For example, when $\tau f_s < 1$, a full cosine cycle is not represented in the metric value.  Consequently, $\mathrm{mean}(E_{\hat{H}})$ does not accurately estimate $\alpha^2$ and the simplification in $(\ref{eq:metric})$ breaks down and yields an additive constant.  Signal processing in other parts of the transceiver, including band-limiting filters, signal conversion, analog circuits, etc. have non-ideal impulse responses.  Subsequently, phase offsets are present.

It should also be noted that as $\alpha$ and $\beta$ approach each other in value, simplifying assumptions of the mean-normalized channel energy in (\ref{eq:metric}) also break down, leading to distortions of the sinusoidal form.  Hence, $\alpha$ and $\beta$ should be significantly different, as assumed by the current system model.  

Unfortunately, we still need to ensure that the power differential between $\alpha$ and $\beta$ does not exceed the maximum resolvable SNR of the receiver.  To address this, the proposed RadCom system will need methods for adjusting $\alpha$ to maximize dynamic range.  For example, the proposed system may employ \textcolor{black}{successive interference cancellation or} analog self-interference cancellation techniques, as described in \cite{Riihonen2012}.  This cancellation should be achievable since empirical results show that a difference of $30-40~\mathrm{dB}$ should be acceptable for small targets at large distances (full duplex transceivers, for example, require much more stringent direct path isolation \cite{Bharadia2013}). {\rt As a result, the need for adjustment of $\alpha$ is an additional feature in RadCom beyond what is natively implemented in a standard IEEE 802.11a/g/p receiver. }
  
\begin{figure}
  \centering
\includegraphics[width=2.5in]{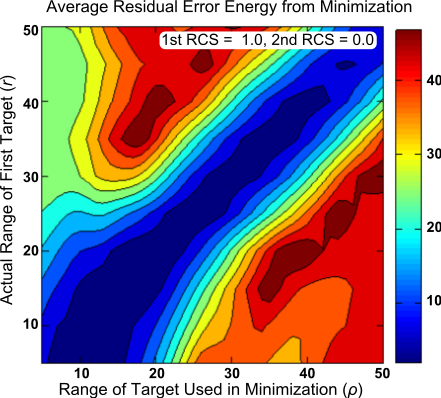}
  \caption{Residual error in cosine model from (\ref{eq:metric}) as a function of actual target range and the range parameter ($\rho$) used during the minimization procedure in Section \ref{sec:ranging1} when a single target is present with RCS $\sigma=1.0$ m$^2$.  No second target is present, which is addressed by setting the RCS of the second target to $0.0$ m$^2$.}
  \label{fig:one_target_residual}
\end{figure}

Consider Fig. \ref{fig:one_target_residual}, which shows the mean residual error for an IEEE 802.11 target detection at $5.89$ GHz with a single target with RCS $\sigma=1$ m$^2$ as a function of the actual target range and the range parameter used for brute-force minimization.
The minimization residual error is very small when the actual target range matches the range parameter since the model used by the minimization algorithm is true.  This verifies that the error function in (\ref{eq:error_func}) is minimized when the range parameter corresponds the actual target range, enabling single-target range estimation through brute-force optimization.

\begin{figure}
  \centering
  \includegraphics[width=2.5in]{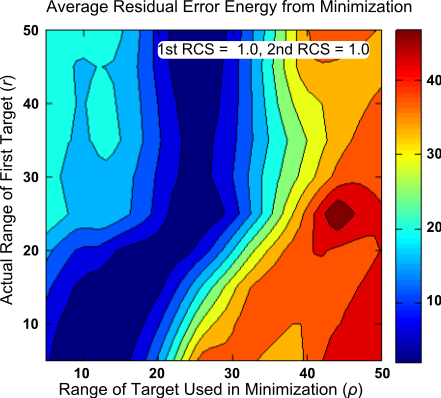}
  \caption{Residual error in cosine model from (\ref{eq:metric}) as a function of first target range and the range parameter ($\rho$) used during the minimization procedure in Section \ref{sec:ranging1} when two targets are present, both with RCS $\sigma=1.0$.  The second target is fixed at a range of $25$ m.}
  \label{fig:two_target_residual}
\end{figure}

Next, consider Fig. \ref{fig:two_target_residual}, which shows the mean residual error of the optimization with two targets present.
This plot shows us that when the first target has a range substantially larger than $25$ m, the fixed location of the second target, the minimization algorithm has a low residual error for target ranges at $25$ m.  The effect is that the proposed ranging algorithm will always focus on the stronger target.  When the two target reflections are of similar strength (around $25$ m), the target range estimate is distorted and detection is somewhat compromised.

\section{Simulation Results}
\label{sec:sims}
This section quantifies the performance of IEEE 802.11 ranging through simulation of the 2-path link model with the proposed minimization procedure in Section \ref{sec:ranging1}.  The summary of parameters of the simulations in this section are listed in Table \ref{table:simpar}.
\begin{table}[t!]
  \caption{Simulation parameters to test ranging feasibility.}
\begin{center}
  \begin{tabular}{|l|l|}
    \hline
    \textbf{Simulation Parameter} & \textbf{Value(s)} \\
    \hline Physical layer & IEEE 802.11a/g/p \\
    \hline Center frequency & $5.89~\mathrm{GHz}$ \\
    \hline Spectrum bandwidth (BW) & $\{10,20\}~\mathrm{MHz}$ \\
    \hline PHY sample rate ($f_s$) & $4\times$ BW \\
    \hline Channel sample rate ($f_s$) & $100\times$ BW \\
    \hline Spectral mask & IEEE 802.11p \\
    \hline Target 1 range ($r$) & $\{5,10,\dots,50\}~\mathrm{m}$ \\
    \hline Target 2 range ($r$) & $25~\mathrm{m}$ \\
    \hline $\mathcal{S}_A$ & $\{-1,-0.5,0,0.5,1\}$ \\
    \hline $\mathcal{S}_C$ & $\{0,\pi/16,\dots,15\pi/16\}$ \\
    \hline $\mathcal{S}_{\rho}$ & $\{5,6,\dots,50\}~\mathrm{m}$ \\
    \hline Channel model & Radar range equation \\
    \hline RCS ($\sigma$) & $\{0.01,0.10,1.00\}~\mathrm{m}^2$ \\
    \hline Noise figure & $5~\mathrm{dB}$ \\ 
    \hline $\epsilon_t$ & $25$ \\
    \hline Path phase & $\sim$ Uniform \\
    \hline Thermal noise & $\sim$ Complex Gaussian \\
    \hline Monte Carlo iterations & 5000 \\
    \hline Transmit power & $20~\mathrm{dBm}$ \\
    \hline Direct path range & $0.1~\mathrm{m}$ \\
    \hline $G_1^{(\text{TX})}, G_1^{(\text{RX})}$ & $0~\mathrm{dBi}$ \\
    \hline $G_2^{(\text{TX})}, G_2^{(\text{RX})}$ & $15~\mathrm{dBi}$ \\
    \hline $F$ & $-70~\mathrm{dB}$ \\
    \hline Channel estimation & Least squares on LTF\\
    \hline
  \end{tabular}
  \label{table:simpar}
  \end{center}
\end{table}
In our simulation, we used the IEEE 802.11p long training field (LTF) to determine receive channel estimates for each target range $r$ defined in Table \ref{table:simpar}.  Using the receive channel estimates determined from each target range $r$, we implemented the minimization algorithm described in Section \ref{sec:ranging1}, using $\mathcal{S}_A$, $\mathcal{S}_C$, $\mathcal{S}_{\rho}$ as specified in Table \ref{table:simpar}.  For each $r$, we found the corresponding range estimate $\rho_{\mathrm{min}}$ returned by the minimization algorithm and determined its root mean square (RMS) error relative to the actual target range $r$.

\begin{figure}
  \begin{center}
    \includegraphics[width=3.25in]{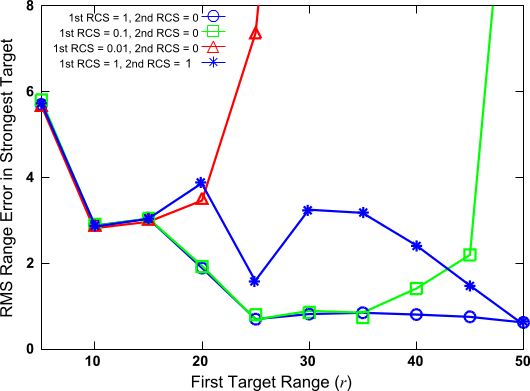}
    \caption{RMS range error when minimization procedure in Section \ref{sec:ranging1} is used on IEEE 802.11 packets in a 10 MHz channel with two targets and variable RCS values. The first target has variable range ($5-50~\mathrm{m}$) and the second target is fixed ($25~\mathrm{m}$).}
    \label{fig:range_error_10_mhz}
  \end{center}
\end{figure}

Fig. \ref{fig:range_error_10_mhz} shows the RMS error of the range estimate (compared to the larger energy target) with IEEE 802.11p and 10 MHz spectral bandwidth using the procedure in Section \ref{sec:ranging1} as a function of target range, for various target RCS values.  In the first three plot lines on this figure, only one RCS value is included (2nd target $\sigma= 0$).  The last plot line reflects two targets with RCS $\sigma = 1.0~\mathrm{m}^2$: the first with variable range and the second target with fixed range at $25~\mathrm{m}$.
This figure shows that strong targets ($\sigma=1.0~\mathrm{m}^2$), in the absence of other targets, can be accurately estimated at distances greater than $15~\mathrm{m}$. Below $15~\mathrm{m}$, the cosine term in (\ref{eq:metric}) does not complete one full cycle.  Consequently, mean offsets are magnified, resulting in occasionally poor estimates.  For weaker targets ($\sigma \leq 0.1 ~\mathrm{m}^2$), range estimates cannot occur accurately at $50~\mathrm{m}$, and are unreliable outside different range boundaries ($45~\mathrm{m}$ for $\sigma=0.1$ and $35~\mathrm{m}$ for $\sigma=0.01~\mathrm{m}^2$).  Fortunately most vehicular targets are significantly greater than $\sigma=1.0~\mathrm{m}^2$.  With two targets, performance is slightly degraded, especially when the two targets have similar, but unequal strength (first target range near $25$ m).  Note that the jaggedness of the RMS range error across different range targets is mostly due to artifacts of the brute-force algorithm as seen from Fig. \ref{fig:verify_10mhz} and \ref{fig:verify_20mhz} and may be improved with more complex algorithms.

\begin{figure}
  \begin{center}
   \includegraphics[width=3.5in]{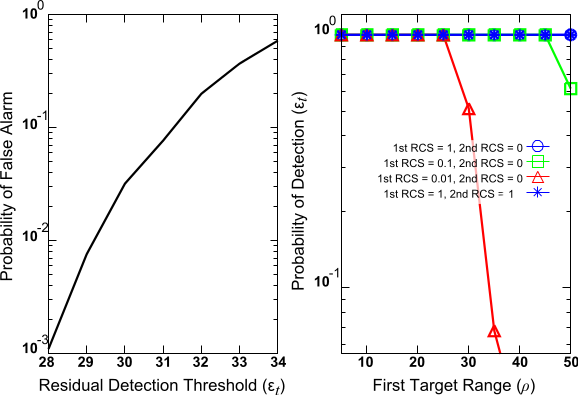}
    \caption{Probability of false alarm (a) and probability of detection (b) when minimization procedure in Section \ref{sec:ranging1} is used on IEEE 802.11 packets in a $10 ~\mathrm{MHz}$ channel with two targets and variable RCS values. The first target has variable range ($5-50~\mathrm{m}$) and the second target is fixed ($25~\mathrm{m}$).}
    \label{fig:pd_fa}
  \end{center}
\end{figure}
 
Fig. \ref{fig:range_error_10_mhz} does not consider detection probability, $p_d$, the probability of false alarm, $p_{fa}$, or the residual error threshold, $\epsilon_t$ (target detection when residual error exceeds ($\epsilon_t$).  To determine the best $\epsilon_t$, we must consider the probability of false alarm, or the probability of detection.  If we assume that ranging estimates will be provided every $50~\mathrm{ms}$ and require that the expected number of false alarm detections $<1/\mathrm{hour}$, this implies $p_{fa}$ is at worst $1\times10^{-5}$.  Extrapolating from Fig. \ref{fig:pd_fa}a, we say that achieving an expected number of false alarm detections $<1/\mathrm{hour}$ requires residual error threshold $\epsilon_t\leq25$.  
We also plot $p_d$ in Fig. \ref{fig:pd_fa}b with $\epsilon_t = 25$.  We note that, as the target range increases and the target RCS decreases, $p_d$ does as well.  Note that there is a region where $p_d$ is reliable, but the range estimate is not.  This suggests that a more conservative $\epsilon_t$ might be desirable to prevent target detection when range estimates are inaccurate.  In the presence of two targets, detection probability is consistent, despite the loss in range precision.  This further motivates a more conservative value for $\epsilon_t$.

The performance shown in Fig. \ref{fig:range_error_10_mhz} is not meter-level accurate.  As mentioned above, for ranges less than $15~\mathrm{m}$, the cosine term in (\ref{eq:metric}) does not complete one full cycle within the mean-normalized channel energy, degrading the performance for the optimization algorithm.  In this case, alternative approaches to range estimation are necessary for accurate detection at close ranges.  These concepts fall outside the scope of this initial study.

To understand the effect of algorithm performance degradation at close ranges as a function of bandwidth, we increase the bandwidth to $20 ~\mathrm{MHz}$. {\rt This is readily supported in IEEE 802.11a/g but is also possible in Channels 175 and 181 in IEEE 802.11p \cite{Kenney2011}.} Note that at ranges less than $5$ m, a $20~\mathrm{MHz}$ channel suffers from performance degradation, i.e., the cosine term in (\ref{eq:metric}) does not complete one full cycle within the mean-normalized channel energy.  We omit ranges less than $5$ m from our detection environment because this performance is acceptable for various applications of vehicular radar, such as mid and long-range radar.  As mentioned before, other range estimation techniques may be employed to improve performance at short ranges, which we reserve for future work.

\begin{figure}
  \begin{center}
    \includegraphics[width=3.25in]{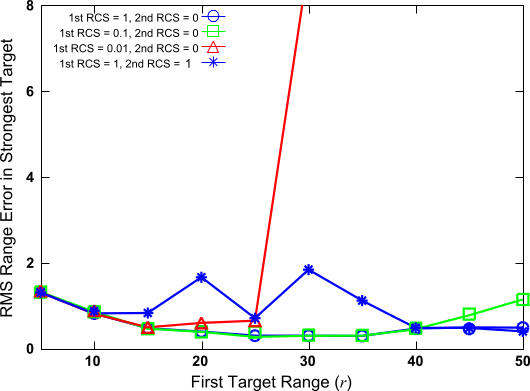}
    \caption{RMS range error when minimization procedure in Section \ref{sec:ranging1} is used on IEEE 802.11 packets in a 20 MHz channel with two targets and variable RCS values. The first target has variable range ($5-50~\mathrm{m}$) and the second target is fixed ($25~\mathrm{m}$).}
    \label{fig:range_error_20_mhz}
  \end{center}
\end{figure}

We simulated the target scenarios from Fig. \ref{fig:range_error_10_mhz} with $20~\mathrm{MHz}$ IEEE 802.11a/g channels in Fig. \ref{fig:range_error_20_mhz}  Note that probability of detection and false alarm did not change significantly, and have not been plotted to save space. 
Fig. \ref{fig:range_error_20_mhz} clearly shows that, in the simulated environment of this paper, our proposed ranging methodology provides sufficient accuracy for vehicular environments.  In single-target environments, meter-level accuracy is achieved at $20 ~\mathrm{MHz}$ when the RCS is sufficiently large.  At a RCS $\sigma = 0.01$m, we see that detection breaks down at 25m.  For much larger vehicular targets, it is possible to achieve meter-level accuracy up to a maximum range of $240$ m with $10 ~\mathrm{MHz}$ of bandwidth based on cyclic prefix duration of the IEEE 802.11p LTF ($120$ m with $20 ~\mathrm{MHz}$ of bandwidth).  Although multi-target effects still noticeably impact system accuracy, the effects are lower than with a $10 ~\mathrm{MHz}$ channel.  Based on these results, we motivate the use of a $20 ~\mathrm{MHz}$ channel for our system implementation.

\section{RadCom Measurement Platform}
\label{sec:platform}

\begin{figure}
  \centering
 \includegraphics[width=3.0in]{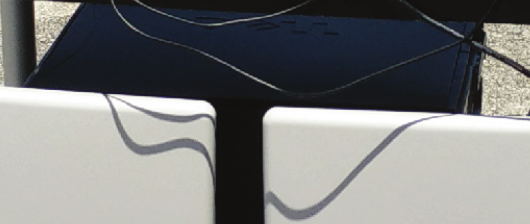}
  \caption{IEEE 802.11 RadCom link setup for measurements.  During measurements the antennas were not as close as in the configuration of this figure.  These antennas were shifted away from each other such that their edges were separated by a minimum of $0.5~m$.}
  \label{fig:setup}
\end{figure}

This section describes the hardware and software platform used to develop our RadCom prototype.  We implemented our algorithm on a host desktop machine and connected it to a software radio transceiver.  By attaching two powerful directional antennas to the transceiver, we were able to extract IEEE 802.11 OFDM channel estimates, which were processed locally through the host machine.  Antennas with high gain and directionality enabled us to blackuce the effect of non-target multipath components within the vehicular environment.  The host machine reported ranging estimates to the user, which were used to demonstrate the performance of the RadCom prototype. An illustration of the physical prototype is provided in Fig.~\ref{fig:setup}. 

To complete over-the-air validation of ranging performance through IEEE 802.11 channel estimates, we leveraged a powerful and configurable IEEE 802.11a/g MAC+PHY LabVIEW FPGA \rt{(field programmable gate array)} 2014 implementation on the National Instruments (NI) Universal Software Radio Peripheral (USRP) Reconfigurable I/O (RIO), provided through the NI lead user program.  This implementation allows a host processing device, i.e., PC computer with LabVIEW host software, to extract channel estimates through a direct memory access (DMA) first-in, first-out (FIFO) buffer on the USRP RIO physically connected to the host through a PCIe interface.  Channel estimates are associated with individual packets which are received correctly (cyclic blackundancy check (CRC) passes).     

Because the channel estimates were already available in the reference lead user design, we did not need to modify any of the LabVIEW FPGA source or recompile the FPGA design.  Instead, we implemented the algorithm from Section \ref{sec:ranging} on the host machine in LabVIEW by creating a new virtual instrument.  This algorithm determined the mean-normalized channel energy from a single received IEEE 802.11 channel estimate.  Using three loops to iterate over the target range candidate, mean offset, and phase offset through a pblackefined set of parameters, the closest-target range estimate was determined through brute-force minimization.  We defined our working set of parameters to be the same as the ones used in our simulations, that is, $\mathcal{S}_A = \{-1, -0.5, 0, 0.5, 1\}$, $\mathcal{S}_C = \{0, \pi/16, \ldots, 15\pi/16\}$, and $\mathcal{S}_{\rho} = \{5, 6, \ldots, 50\}$m.

With this implementation we were able to compute ranging estimates on intervals of 150 msec through beacon evaluation. Future implementations will desire the use security-enabled traffic, but traffic type was not important for the purposes of this feasibility study.  With a FPGA implementation of the algorithm in Section \ref{sec:ranging}, we could have pushed this update interval much lower, but this was not necessary for feasibility testing.

\begin{figure}
\centering
  \includegraphics[width=3in]{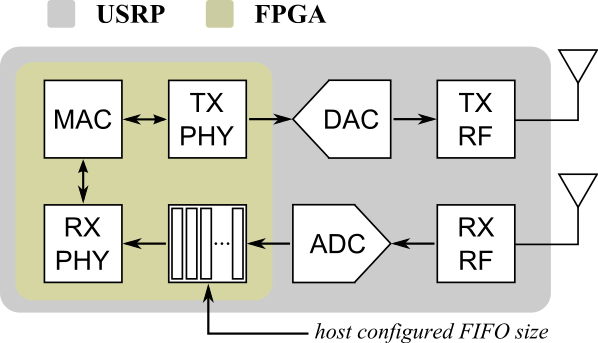}
  \caption{Block diagram of IEEE 802.11a/g implementation as it relates to functionality roles and their allocation on the FPGA-controlled NI USRP RIO platform.}
  \label{fig:80211_block_diagram}
\end{figure}

In standard IEEE 802.11 communication links, the transmitter and receiver do not operate simultaneously.  This is also the case our design.  Consider Fig. \ref{fig:80211_block_diagram}, which shows how different pieces of the 802.11 implementation are allocated on different platform resources.
The key observation from this figure is that a FIFO of configurable size is placed between the analog to digital converter (ADC) and the receiver PHY.  If the FIFO size is large enough, a sufficient delay is created between the transmit and receive paths of the same device.  In this case the IEEE 802.11a/g implementation can send and receive its own packet.  This step is critical for radar processing.

\begin{figure*}
\centering
    \includegraphics[width=\textwidth]{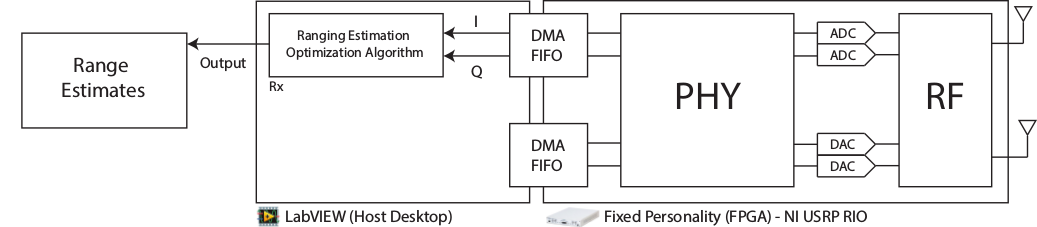}
    \caption{Block diagram of I/O relationship of measurement platform.  Antennas are attached to the NI USRP RIO, which communicates with LabVIEW on the host desktop via DMA FIFO links.}
    \label{fig:software_block}
\end{figure*}
The NI USRP-2953R has a tunable center frequency from 1.2 GHz to 6 GHz with a 40 MHz per channel real-time bandwidth and supports an 800 MB/s connection to the host.  The bandwidth and data rate provided by the USRP RIO are more than sufficient for the purposes of this study. Figure \ref{fig:software_block} illustrates the I/O relationship between the NI USRP RIO and software.  \textcolor{black}{For more details on the time and frequency characteristics of the 802.11 waveforms, please consult \cite{Waveforms80211ac}.}

\begin{table}[t!]
  \caption{Hardware specifications for our prototype.}
\begin{center}
  \begin{tabular}{| >{\raggedleft\let\newline\\\arraybackslash\hspace{0pt}}m{2cm} | >{\raggedleft\let\newline\\\arraybackslash\hspace{0pt}}m{2cm} | >{\raggedleft\let\newline\\\arraybackslash\hspace{0pt}}m{3.5cm}|}
    \hline \textbf{Hardware} & \textbf{Specifications} & \textbf{Value(s)} \\
    \hline \multirow{16}{2cm}{\textbf{NI USRP-2953R}} 	
		& Frequency & $1.2-6.0 ~\mathrm{GHz}$ \\ \cline{2-3}
		& Frequency Step & $<1~\mathrm{kHz}$\\ \cline{2-3}
		& Maximum Output Power & $17~\mathrm{dBm}$ to $20 ~\mathrm{dBm}$ @ $1.2-3.5 ~\mathrm{GHz}$, $7 ~\mathrm{dBm}$ to $15 ~\mathrm{dBm}$ @ $3.5-6 ~\mathrm{GHz}$\\ \cline{2-3}
		& Frequency Accuracy & $25$ ppb (unlocked)\\ \cline{2-3}
		& Maximum Instantaneous Real-Time Bandwidth & $40 ~\mathrm{MHz}$ per Channel\\ \cline{2-3}
		& Digital-to-Analog Converter & Sample Rate $400$ MS/s, Resolution $16$ bit, Spurious-free Dynamic Range $80 ~\mathrm{dB}$\\\cline{2-3}
    \hline \multirow{14}{2cm}{\textbf{L-COM $23 ~ \mathrm{dBi}$ Broadband Patch Antennas}}
		& Frequency Range & $4750-5850 ~\mathrm{MHz}$ \\ \cline{2-3}
		& L-COM Item \# & HG4958-23P \\ \cline{2-3} 
		& Dimensions & $315$ x $315$ x $25$ mm \\ \cline{2-3}
		& Gain & $20 ~\mathrm{dBi}$ @ $4.9 ~\mathrm{GHz}$/$23 ~\mathrm{dBi}$ @ $5.8 ~\mathrm{GHz}$ \\\cline{2-3}
		& Polarization & Vertical or Horizontal \\\cline{2-3}
		& Horizontal Beam Width & $11^\circ$ \\\cline{2-3} 
		& Vertical Beam Width & $11^\circ$ \\ \cline{2-3}
    \hline
  \end{tabular}
  \label{table:specs}
  \end{center}
\end{table}

We chose a lower center frequency of $4.89~\mathrm{GHz}$ for our study instead of $5.9~\mathrm{GHz}$ due to the limited power output of the transceiver hardware.  All of the RF components in our measurement setup are compatible with this frequency.
\begin{figure}
  \begin{center}
    \includegraphics[width=3.5in]{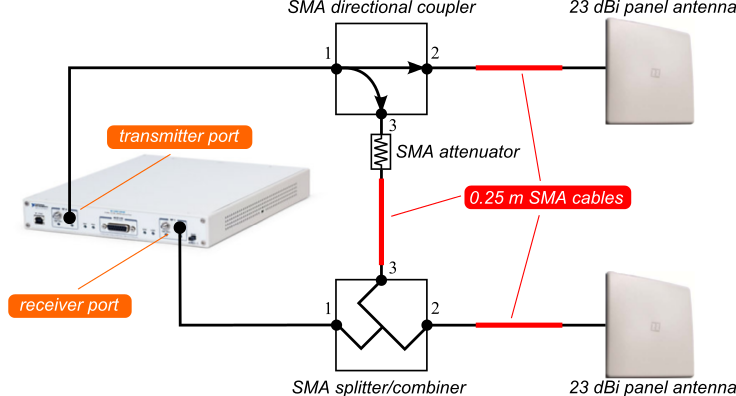}
    \caption{Block diagram description of RF components in the measurements campaign.  The reflected path (from the antennas) uses $0.5$ m of SMA cable, but the direct path (from coupled port to combiner port) only uses $0.25$ m of SMA cable, which artificially adds $0.125$ m to the target range (results have been adjusted accordingly).}
    \label{fig:microwave_components}
  \end{center}
\end{figure}
Focusing on targets and isolating non-target multipath components requires high directionality and gain from antennas.  For this purpose, we attached L-COM $23 ~\mathrm{dBi}$ HG4958-23P Broadband Patch Antennas to the transmitter and receiver SMA ports on the USRP RIO through a microwave component network.  To optimize the performance of our implementation, we chose patch antennas with especially high gain.  This ensublack that low energy signals and maximum detectable range would primarily be due to our system model and algorithm and not due to hardware limitations.  In practice, smaller antennas with lower gain should be used to blackuce cost while maintaining high performance.  Future work should study the effect of lower gain antennas on the RadCom platform.

The RF component configuration attached to the NI USRP RIO device is depicted in Fig. \ref{fig:microwave_components}. The transmitter SMA port output feeds into the input port (coupler port $1$) of a $3$-port Narda 4053-30 directional coupler through a variable-resistance RF attenuator at the output port (coupler port $3$) followed by a $0.25$ m SMA cable and an input port (combiner port $3$) of a $3$-port RF combiner.  The directional coupler was chosen over an RF splitter/combiner to isolate the direct path from the signals captublack on the transmit antenna.  Note that we use the RF attenuator value to scale the direct path appropriately in relation to the reflected path of our two-path model.  The strength of the reflected path varies as target distance varies, so different attenuator values were used during the experiment. The coupled port (coupler port $3$) serves as the direct path source and the output port (coupler port $2$) serves as the antenna source. The output port of the directional coupler (coupler port $2$) feeds into the transmit antenna through another $0.25$ m SMA cable.  The receive antenna feeds into the remaining combiner input port (combiner port $2$) through a $0.25$ m SMA cable, and the combiner output port (combiner port $1$) feeds directly into the receiver SMA port.  The approximate scattering matrix of the Narda directional coupler and the RF combiner are
\begin{eqnarray}
\mathbf{S}_{\mathrm{coup}} = \left[ \begin{array}{ccc} -20 & 0 & -50 \\ 0 & -20 & -50 \\ -30 & -50 & -20 \end{array} \right]\\
\mathbf{S}_{\mathrm{comb}} = \left[ \begin{array}{ccc} -20 & -6 & -6 \\ -6 & -20 & -6 \\ -6 & -6 & -20 \end{array} \right]\nonumber
\end{eqnarray}
where the values in the scattering matrix are defined in decibels.   

A 1000 W DC-to-AC converter was attached to the battery of a mid-sized vehicle to support the power requirements of the NI USRP RIO platform and the Dell Desktop Precision T5500 host PC. The antennas were attached to the USRP and placed roughly one-third of a meter off the ground as illustrated in Fig. \ref{fig:setup}.  During measurements the outside edges of the antennas were separated by a minimum of $0.5$ m, which is larger than depicted in this figure.

During preliminary studies, we observed an irregular effect in the channel impulse response when the antennas were located too close to the ground.  We determined that the antennas should be located sufficiently above the ground on the vehicle in implementation.  Based on existing vehicular design, the ideal locations for the antennas would be in close proximity to the vehicle headlights.

\section{Measurement Results}
\label{sec:scenarios}
\begin{table}[t!]
\begin{center}
  \caption{Testing parameters}
  \label{table:params}
  \begin{tabular}{|l|l|}
    \hline
    \textbf{Parameter} & \textbf{Value(s)} \\
    \hline Center Frequency & $4.89 ~\mathrm{GHz}$\\
    \hline TX Power& $10 ~\mathrm{dBm}$\\
    \hline Operation Mode& RF, with delay\\
    \hline Modulation Scheme & QPSK, rate $3/4$ \\
    \hline Transmission Mode & $20 ~\mathrm{MHz}$ (IEEE 802.11a/g) \\
    \hline
  \end{tabular}
  \end{center}
\end{table}

\begin{figure}
  \begin{center}
    \includegraphics[width=3.5in]{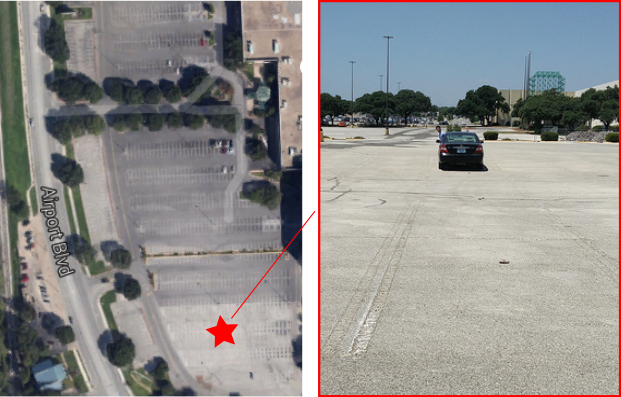}
    \caption{Left: Map of measurement location in a vacant parking lot at Highland Mall, Austin, TX (retrieved from Google Maps).  Right: Image of measurement location with vehicle target (2002 Toyota Camry).}
    \label{fig:measurement_location}
  \end{center}
\end{figure}

The measurements reported in this paper were conducted in a vacant parking lot of Austin Community College in North-Central Austin.  This site was chosen to isolate a vehicle target in a forward collision scenario (using the vehicle rear end as the radar cross section).  The measurement location is visualized in Fig. \ref{fig:measurement_location}.

Both the transmit and receive antenna are pointed towards the target, the rear surface of a 2002 Toyota Camry.  The RadCom platform from Fig. \ref{fig:setup} was initially positioned at a distance of $30$ m from the target.  To collect measurements, we transmitted IEEE 802.11 beacon messages with the configuration in Table \ref{table:params} every 250 milliseconds.  After a sufficient number of channel estimates were obtained at $30$ m from the target, the RadCom platform moved $5$ m in a straight line toward the target.  After a sufficient number of channel estimates were obtained at $25$ m distance from the target, the RadCom platform moved another $5$ m towards the target.  This process continued until measurements were obtained at all $5$ m target distance increments.

To ensure that the energy of the direct path did not overwhelm the reflected path, the RF attenuation in the direct path was adjusted between $0-30$ dB, depending on the distance.  These values of attenuation were empirically determined and allow the target reflection path energy to remain comparable to the direct path's energy, which ensured that the target reflection path component was accurately resolved in the channel estimate.  The direct path also experienced $30$ dB attenuation due to the coupling effect and $6$ dB attenuation in the combiner for a total of $6-66$ dB attenuation in the direct path.  The RCS equation predicts $\approx 39-70$ dB attenuation for the target reflection path when $23$ dBi antennas are used with a $1$ m$^2$ RCS target.  Hence, the empirical observation seems reasonable from simple link budget analysis.

During data capture we periodically measured the frequency domain channel response of the direct path by terminating the antenna path.  While the energy of the channel impulse response of the direct path was linear, there was often a gradual linear slope.  This slope resulted in a fluctuation of a few dB from the left edge subcarrier to right edge subcarrier.  Essentially, the direct path could not accurately be modeled by a single tap, but instead required two taps separated by a very small delay (less than the sample period).  The non-ideal direct path is likely due to internal feedthrough or internal reflections in microwave components.  A product-ready solution would not have these effects.  When the antenna path was activated for data capture, the presence of this second tap in the direct path (and third tap overall) degraded the performance of the ranging algorithm.  
Fortunately, a simple procedure was discovered to calibrate the data and remove the impairment caused by the non-ideal direct path.  For example, consider that a three tap channel can be modeled in the frequency domain channel response with
\begin{equation}
        H[m] = \alpha + \beta e^{-j \left( 2\pi m \Delta \tau + \theta \right) } + \gamma e^{-j \left( 2\pi m \Delta \tau_D + \phi \right) }
\end{equation}
where the last additive term is due to the addition of a second tap in the direct path $\tau_D$ seconds after the first tap represented by $\alpha$.
Next, consider the energy of the frequency domain channel response with this model as 
\begin{align}
E_H[m] & =
 C + \alpha\beta\cos\left( 2\pi m \Delta \tau + \theta \right) + \alpha \gamma \cos\left( 2\pi \Delta \tau_D m + \phi \right) 
\nonumber \\
& \quad \quad +~ \beta\gamma\cos\left( 2\pi m \left(\tau-\tau_D\right) + \theta - \phi \right) \\
    & \approx  C + \alpha\beta\cos\left( 2\pi m \Delta \tau + \theta \right) \nonumber \\
    & \quad \quad  + ~\alpha\gamma\cos\left( 2\pi \Delta \tau_Dm + \phi \right) \label{eq:calibration}
\end{align}
given $C\stackrel{\Delta}{=}\alpha^2+\beta^2+\gamma^2$.  The approximation follows from the assumption that $\beta,\gamma \ll \alpha$.  Hence, the channel energy with the non-ideal term simply includes an additional cosine term.  
Empirically we observed that $\tau_D$ is small compared to the sample time and that the $50$\% coherence bandwidth is larger than the channel bandwidth \cite{Rappaport2002}.  A linear function approximation was sufficient to estimate the slope that resulted from the contribution from the second tap in the direct path.  Therefore, we manually removed any slope in the frequency domain channel impulse response before algorithm processing.  This slope was easily discovered since we were able to analyze its effect over many continuous channel estimates.

The statistics of the range estimates extracted from measurements are presented in Table \ref{table:data}. The range estimate statistics (mean and standard deviation) in terms of measurement distance and number of data points collected when the minimization procedure in Section \ref{sec:ranging} is used on IEEE 802.11 packets in a $20~\mathrm{MHz}$ channel with one target as in scenario from Section \ref{sec:scenarios}.  Post-calibration estimates are presented.
\begin{table}[t!]
\begin{center}
  \caption{Range estimate statistics}
  \label{table:data}
  \begin{tabular}{|l|l|l|l|}
    \hline
    \textbf{Distance} & \textbf{Data Points} & \textbf{Est. Mean} & \textbf{Est. STD}  \\
    \hline $5$ m & 100 & $6.00$ m & $0.00$ m\\
    \hline $10$ m & 100 & $9.96$ m & $0.67$ m\\
    \hline $15$ m & 100 & $14.03$ m & $0.30$ m\\
    \hline $20$ m & 87 & $19.48$ m & $0.50$ m\\
    \hline $25$ m & 100 & $25.96$ m & $0.24$ m\\
    \hline $30$ m & 100 & $29.12$ m & $0.33$ m\\
    \hline
  \end{tabular}
  \end{center}
\end{table}
The root mean square (RMS) error of the range estimate is plotted in Fig. \ref{fig:rangerms}.
\begin{figure}
  \centering
  \includegraphics[width=3.0in]{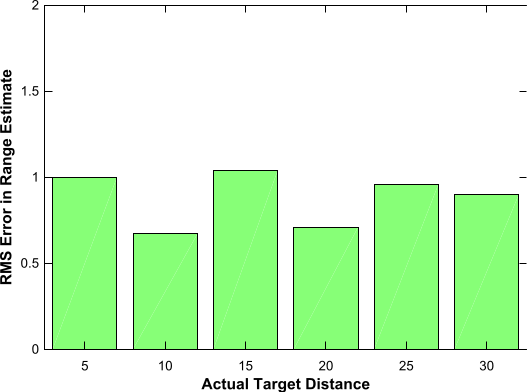}
  \caption{RMS range error when minimization procedure in Section \ref{sec:ranging} is used on IEEE 802.11 packets in a $20 ~\mathrm{MHz}$ channel with one target as in scenario from Section \ref{sec:scenarios}.  The target has variable range from $5$ to $30$m.  Post-calibration estimates are presented.}
  \label{fig:rangerms}
\end{figure}
The results of this study have demonstrated feasibility for IEEE 802.11 radar in a vehicular environment.  Based on our setup from Section \ref{sec:platform}, our system can be implemented on currently existing IEEE 802.11 devices with minimal modification to the physical layer, supporting a secure and extremely cost-effective design.  We suggest that with slight modifications, the results from this study can be extended to a DSRC-based platform with a $10~\mathrm{MHz}$ bandwidth.  Future work will address ways to mitigate the effect of bandwidth reduction in IEEE 802.11 radar.

%

\section{Results and Discussion}
\label{sec:discussion}
The statistics of the range estimates extracted from measurements are presented in Table \ref{table:data}.
\begin{table}[t!]
\begin{center}
  \caption{Range estimate statistics (mean and standard deviation) in terms of measurement distance and number of data points collected when minimization procedure in Section \ref{sec:ranging} is used on IEEE 802.11 packets in a $20~\mathrm{MHz}$ channel with one target as in scenario from Section \ref{sec:scenarios}.  Post-calibration estimates are presented.}
  \label{table:data}
  \begin{tabular}{|l|l|l|l|}
    \hline
    \textbf{Distance} & \textbf{Data Points} & \textbf{Estimate Mean} & \textbf{Estimate STD}  \\
    \hline $5$ m & 100 & $6.00$ m & $0.00$ m\\
    \hline $10$ m & 100 & $9.96$ m & $0.67$ m\\
    \hline $15$ m & 100 & $14.03$ m & $0.30$ m\\
    \hline $20$ m & 87 & $19.48$ m & $0.50$ m\\
    \hline $25$ m & 100 & $25.96$ m & $0.24$ m\\
    \hline $30$ m & 100 & $29.12$ m & $0.33$ m\\
    \hline
  \end{tabular}
  \end{center}
\end{table}
The root mean square (RMS) error of the range estimate is plotted in Fig. \ref{fig:rangerms}.
\begin{figure}
  \centering
  \caption{RMS range error when minimization procedure in Section \ref{sec:ranging} is used on IEEE 802.11 packets in a $20 ~\mathrm{MHz}$ channel with one target as in scenario from Section \ref{sec:scenarios}.  The target has variable range from $5$ to $30$m.  Post-calibration estimates are presented.}
  \label{fig:rangerms}
\end{figure}
The results of this study have demonstrated feasibility for IEEE 802.11 radar in a vehicular environment.  Based on our setup from Section \ref{sec:platform}, our system can be implemented on currently existing IEEE 802.11 devices with minimal modification to the physical layer, supporting a secure and extremely cost-effective design.  We propose a collision avoidance system design that uses this IEEE 802.11 radar in conjunction with current mmWave systems in order to provide an essential layer of security.  In this design, the ranging information received by the IEEE 802.11 radar can be used as a cross-reference to the data received by the mmWave radar, verifying the accuracy of the mmWave ranging and providing blackundancy to the overall system.  By using DSRC, additional security can be implemented by designing the system to verify the communication waveform at the receiver.  Our proposed system addresses the vulnerabilities of existing mmWave systems in a non-intrusive, low-cost, and efficient way, providing confidence that the collision avoidance systems used in automotive transportation today are assublack to be safe and secure.  Despite this promise, substantially more work is needed to improve performance in practical vehicular radar environments.

Our results only consider a single vehicle.  Roadways, however, feature a multitude of vehicle types and physical dimensions.  We do not currently understand how the performance of IEEE 802.11 radar will perform for different vehicles.  Also, it is possible that different reflective surfaces may produce different received signal strengths in the channel impulse response.  Past work at mmWave frequencies has shown that the radar cross section of different vehicles can vary from $16 ~\mathrm{dBsm}$ for 
trucks to $14 ~\mathrm{dBsm}$ for small vehicles to $-1 ~\mathrm{dBsm}$ for motorcycles \cite{Suzuki2000}.

In practical vehicular radar environments there will be many false targets in the channel that can corrupt the performance of our single target algorithm.  It is reasonable to assume that the detection environment has many potential targets of interest. Consider the simple scenario where multiple objects are roughly the same distance away from the source antennas, and have roughly the same RCS. A real-world extension could be a vehicle directly in front of the source and a vehicle in the adjacent lane at the same distance. Even if the channel estimates were retrieved from highly directional antennas, we would not be able to distinguish whether or not the detected target was directly in front or in an adjacent lane without additional modification to the setup described. Thus, the strategies describing single-target detection must be extrapolated to multiple targets with care.

Multiple targets detection with highly varying RCSs may also become a difficult challenge. As described above, the resulting signal strength is dependent on the RCS. If there is a small target near the source antennas and a larger target farther away from the source, then it is possible that target ambiguity could be introduced to the detection scheme based on the reflected signal strength. In practice, this may present challenges when considering motorcycle detection. 

\section{Conclusions}
\label{sec:conclusion}

\rt{In this paper, we evaluated an approach for using IEEE 802.11a/g/p waveforms for automotive radar. We described the approach based on a two-path channel model, proposed an algorithm for extracting the range parameters, and evaluated its performance using simulations and measurements. We focused our attention on IEEE 802.11p, which is the PHY of DSRC, considering both the recommended $10$ MHz of bandwidth and also the optional $20$ MHz of bandwidth. We found that our approach could achieve $1$ m accuracy for a single target even with $10$ MHz of bandwidth, but required $20$ MHz to achieve similar performance in the two-target scenario. Future work should consider improving performance for medium and short range applications, which requires better cancellation of the direct path, coherent processing across multiple IEEE 802.11p packets for Doppler estimation, and the benefits of multiple antennas as found in IEEE 802.1n/ac for better separation of targets in azimuth. }

\bibliographystyle{IEEEbib}
\bibliography{refs}

\begin{thebibliography}{10}

\bibitem{Naughton2017}
K.~Naughton and M.~Bergen,
\newblock ``Alphabet's waymo cuts cost of key self-driving sensor by {90\%},''
  2017.

\bibitem{SmithPeter2000}
C.~Smithpeter, R.~Nellums, S.~Lebien, and G.~Studor,
\newblock ``A miniature, high-resolution laser radar operating at video
  rates,''
\newblock in {\em Laser Radar Technology and Applications}, 2000, pp. 279--286.

\bibitem{Carullo2001}
A.~Carullo and M.~Parvis,
\newblock ``An ultrasonic sensor for distance measurement in automotive
  applications,''
\newblock {\em IEEE Sensors Journal}, vol. 1, no. 2, pp. 143--147, 2001.

\bibitem{Klotz2000}
Michael Klotz and Hermann Rohling,
\newblock ``24 {GHz} radar sensors for automotive applications,''
\newblock in {\em IEEE International Conference on Microwaves, Radar and
  Wireless Communications}, 2000, vol.~1, pp. 359--362.

\bibitem{Kenney2011}
J.~Kenney,
\newblock ``Dedicated short-range communications ({DSRC}) standards in the
  {United States},''
\newblock {\em Proceedings of the IEEE}, vol. 99, no. 7, pp. 1162--1182, 2011.

\bibitem{Hasch2012}
J.~Hasch, E.~Topak, R.~Schnabel, T.~Zwick, R.~Weigel, and C.~Waldschmidt,
\newblock ``Millimeter-wave technology for automotive radar sensors in the 77
  {GHz} frequency band,''
\newblock {\em IEEE Transactions on Microwave Theory and Techniques}, vol. 60,
  no. 3, pp. 845--860, 2012.

\bibitem{Rappaport2014}
T.~Rappaport, R.~{Heath, Jr.}, R.~Daniels, and J.~Murdock,
\newblock {\em Millimeter Wave Wireless Communications},
\newblock Pearson Education, 2014.

\bibitem{Cerretelli2007}
M.~Cerretelli and G.~Gentili,
\newblock ``Progress in compact multifunction automotive antennas,''
\newblock in {\em Proceedings of the IEEE International Conference on
  Electromagnetics in Advanced Applications}, 2007, pp. 93--96.

\bibitem{Hubaux2004}
J.~Hubaux, S.~Capkun, and J.~Luo,
\newblock ``The security and privacy of smart vehicles,''
\newblock {\em IEEE Security \& Privacy Magazine}, vol. 2, pp. 49--55, 2004.

\bibitem{Chauhan2014}
R.~Chauhan,
\newblock ``A platform for false data injection in frequency modulated
  continuous wave radar,''
\newblock M.S. thesis, Utah State University, 2014.

\bibitem{Yeh2016}
E.~Yeh, J.~Choi, N.~Prelcic, C.~Bhat, and R.~{Heath, Jr.},
\newblock ``Security in automotive radar and vehicular networks,''
\newblock Submitted to Microwave Journal, 2016.

\bibitem{Brooker2007}
G.~Brooker,
\newblock ``Mutual interference of millimeter-wave radar systems,''
\newblock {\em IEEE Transactions on Electromagnetic Compatibility}, vol. 49,
  pp. 170--181, 2007.

\bibitem{Sturm2011}
Christian Sturm and Werner Wiesbeck,
\newblock ``Waveform design and signal processing aspects for fusion of
  wireless communications and radar sensing,''
\newblock {\em Proceedings of the IEEE}, vol. 99, no. 7, pp. 1236--1259, 2011.

\bibitem{TTT2015}
``Toyota to introduce world's first {DSRC}-based {V2X} system in production
  cars,'' \url{http://www.traffictechnologytoday.com/news.php?NewsID=73445},
  October 2015.

\bibitem{USDOT2015}
``Teleconference transcript,''
  \url{http://www.its.dot.gov/itspac/november2015/ITSPAC_2015-11-13_Transcript.pdf},
  2015.

\bibitem{Kihei2015}
B.~Kihei, J.~Copeland, and Y.~Chang,
\newblock ``Design considerations for vehicle-to-vehicle {IEEE 802.11p} radar
  in collision avoidance,''
\newblock in {\em Proceedings of the IEEE Global Communications Conference},
  2015, pp. 1--7.

\bibitem{Rohling2001}
H.~Rohling and M.~Meinecke,
\newblock ``Waveform design principles for automotive radar systems,''
\newblock in {\em Proceedings of the CIE International Conference on Radar},
  2001, pp. 1--4.

\bibitem{Van2011}
R.~{Van Nee},
\newblock ``Breaking the gigabit-per-second barrier with 802.11ac,''
\newblock {\em IEEE Wireless Communications}, vol. 18, no. 2, pp. 4, 2011.

\bibitem{Colone2012}
F.~Colone, P.~Falcone, C.~Bongioanni, and P.~Lombardo,
\newblock ``{WiFi}-based passive bistatic radar: data processing schemes and
  experimental results,''
\newblock {\em IEEE Transactions on Aerospace and Electronic Systems}, vol. 48,
  no. 2, pp. 1061--1079, 2012.

\bibitem{Falcone2012}
P.~Falcone, F.~Colone, and P.~Lombardo,
\newblock ``Potentialities and challenges of {WiFi}-based passive radar,''
\newblock {\em IEEE Aerospace and Electronic Systems Magazine}, vol. 27, no.
  11, pp. 15--26, 2012.

\bibitem{Chetty2012}
K.~Chetty, G.~Smith, and K.~Woodbridge,
\newblock ``Through-the-wall sensing of personnel using passive bistatic {WiFi}
  radar at standoff distances,''
\newblock {\em IEEE Transactions on Geoscience and Remote Sensing}, vol. 50,
  no. 4, pp. 1218--1226, 2012.

\bibitem{Maechler2012}
P.~Maechler, N.~Felber, and H.~Kaeslin,
\newblock ``Compressive sensing for {WiFi}-based passive bistatic radar,''
\newblock in {\em Proceedings of the IEEE Europeans Signal Processing
  Conference}, 2012, pp. 1444--1448.

\bibitem{Ivashko2014}
I.~Ivashko, O.~Krasnov, and A.~Yarovoy,
\newblock ``Receivers topology optimization of the combined active and
  {WiFi}-based passive radar network,''
\newblock in {\em Proceedings of the IEEE European Radar Conference}, 2014, pp.
  517--520.

\bibitem{Rzewuski2013}
S.~Rzewuski, M.~Wielgo, K.~Kulpa, M.~Malanowski, and J.~Kulpa,
\newblock ``Multistatic passive radar based on {WiFi}-results of the
  experiment,''
\newblock in {\em Proceedings of the IEEE International Conference on Radar},
  2013, pp. 230--234.

\bibitem{Adib2013}
F.~Adib and D.~Katabi,
\newblock ``See through walls with {WiFi!},''
\newblock in {\em Proceedings of the ACM Special Interest Group on Data
  Communication Conference}, 2013.

\bibitem{Letzepis2011}
N.~Letzepis, A.~Grant, P.~Alexander, and D.~Haley,
\newblock ``Joint estimation of multipath parameters from {OFDM} signals in
  mobile channels,''
\newblock in {\em Proceedings of the IEEE Australian Communications Theory
  Workshop}, 2011, pp. 106--111.

\bibitem{Mizui1993}
K.~Mizui, M.~Uchida, and M.~Nakagawa,
\newblock ``Vehicle-to-vehicle communication and ranging system using spread
  spectrum technique,''
\newblock in {\em Proceedings of the IEEE Vehicular Technology Conference},
  1993, pp. 335--338.

\bibitem{Sturm2010}
C.~Sturm, M.~Braun, T.~Zwick, and W.~Wiesbeck,
\newblock ``A multiple target {Doppler} estimation algorithm for {OFDM} based
  intelligent radar systems,''
\newblock in {\em Proceedings of the European Radar Conference}, 2010, pp.
  73--76.

\bibitem{Sit2011radar}
Y.~Sit, C.~Sturm, and T.~Zwick,
\newblock ``Doppler estimation in an {OFDM} joint radar and communication
  system,''
\newblock in {\em Proceedings of the 6th German Microwave Conference}, 2011,
  pp. 1--4.

\bibitem{Braun2009}
M.~Braun, C.~Sturm, A.~Niethammer, and F.~Jondral,
\newblock ``Parametrization of joint {OFDM}-based radar and communication
  systems for vehicular applications,''
\newblock in {\em Proceedings of PIRMC}, 2009, pp. 3020--3024.

\bibitem{Garmatyuk2009}
D.~Garmatyuk, J.~Schuerger, K.~Kauffman, and S.~Spalding,
\newblock ``Wideband {OFDM} system for radar and communications,''
\newblock in {\em Proceedings of the IEEE Radar Conference}, 2009, pp. 1--6.

\bibitem{Crouse2012}
D.~Crouse,
\newblock ``A time-shift model for {OFDM} radar,''
\newblock in {\em Proceedings of the IEEE Radar Conference}, 2012, pp.
  841--846.

\bibitem{Sit2014}
Y.~Sit and T.~Zwick,
\newblock ``{MIMO} {OFDM} radar with communication and interference
  cancellation features,''
\newblock in {\em Proceedings of the IEEE Radar Conference}, 2014, pp. 19--23.

\bibitem{Kashin2013}
V.~Kashin and E.~Mavrychev,
\newblock ``Target velocity estimation in {OFDM} radar based on subspace
  approaches,''
\newblock in {\em Proceedings of the 14th International Radar Symposium}, 2013,
  pp. 19--21.

\bibitem{Sen2011}
S.~Sen and A.~Nehorai,
\newblock ``Sparsity-based multi-target tracking using {OFDM} radar,''
\newblock {\em IEEE Transaction on Signal Processing}, vol. 59, pp. 1902--1906,
  2011.

\bibitem{Reichardt2012}
L.~Reichardt, C.~Sturm, F.~Grunhaupt, and T.~Zwick,
\newblock ``Demonstrating the use of the {IEEE 802.11p} car-to-car
  communication standard for automotive radar,''
\newblock in {\em Proceedings of the 6th European Conference on Antennas and
  Propagation}, 2012, pp. 1576--1580.

\bibitem{Alexander2015}
P.~Alexander, N.~Letzepis, A.~Grant, and D.~Haley,
\newblock ``Estimation of a multipath signal in a wireless communication
  system,'' 2015,
\newblock US Patent 9,083,419.

\bibitem{CohdaWireless2016}
Cohda Wireless,
\newblock ``{GPSless} positioning for {V2X},'' 2016.

\bibitem{Jiang2008}
D.~Jiang and L.~Delgrossi,
\newblock ``{IEEE} 802.11p: Towards an international standard for wireless
  access in vehicular environments,''
\newblock in {\em Proceedings of the IEEE Vehicular Technology Conference},
  2008, pp. 2036--2040.

\bibitem{Sit2011}
Y.~Sit, L.~Reichardt, C.~Sturm, and T.~Zwick,
\newblock ``Extension of the {OFDM} joint radar-communication system for a
  multipath, multiuser scenario,''
\newblock in {\em Proceedings of the IEEE Radar Conference}, 2011, pp.
  718--723.

\bibitem{Kumari2015}
P.~Kumari, N.~Prelcic, and R.~{Heath, Jr.},
\newblock ``Investigating the {IEEE} 802.11ad standard for millimeter wave
  automotive radar,''
\newblock in {\em Proceedings of the IEEE Vehicular Technology Conference},
  2015.

\bibitem{Marchetti2009}
Marchetti N, M.~Rahman, S.~Kumar, and R.~Prasad,
\newblock ``{OFDM}: Principles and challenges,''
\newblock in {\em New Directions in Wireless Communications Research}, pp.
  29--62. Springer, 2009.

\bibitem{Wang2006}
T.~Wang, J.~Proakis, E.~Masry, and J.~Zeidler,
\newblock ``Performance degradation of {OFDM} systems due to {Doppler}
  spreading,''
\newblock {\em IEEE Transactions on Wireless Communications}, vol. 5, pp.
  1422--1432, 2006.

\bibitem{Yucek2005}
T.~Yucek, R.~Tannious, and H.~Arslan,
\newblock ``Doppler spread estimation for wireless {OFDM} systems,''
\newblock in {\em Proceedings of the IEEE/Sarnoff Symposium on Advances in
  Wired and Wireless Communication}, 2005, pp. 233--236.

\bibitem{Stove1992}
A.~Stove,
\newblock ``Linear {FMCW} radar techniques,''
\newblock {\em IEE Proceedings F - Radar and Signal Processing}, vol. 139, pp.
  343--350, 1992.

\bibitem{Cheng2008}
L.~Cheng, B.~Henty, D.~Stancil, and F.~Bai,
\newblock ``Doppler component analysis of the suburban vehicle-to-vehicle
  {DSRC} propagation channel at {5.9 GHz},''
\newblock in {\em Proceedings of the IEEE Radio and Wireless Symposium}, 2008,
  pp. 343--346.

\bibitem{Welsh1967}
P.~Welch,
\newblock ``The use of fast fourier transform for the estimation of power
  spectra: a method based on time averaging over short, modified
  periodograms,''
\newblock {\em IEEE Transactions Audio and Electroacoustic}, vol. 15, no. 1,
  pp. 17--20, 1967.

\bibitem{Gabriel1986}
W.~Gabriel,
\newblock ``Using spectral estimation techinques in adaptive processing antenna
  systems,''
\newblock {\em IEEE Transactions on Antennas and Propagation}, vol. 34, no. 3,
  pp. 291--300, 1986.

\bibitem{Byrnes2000}
C.~Byrnes, T.~Georgiou, and A.~Lindquist,
\newblock ``A new approach to spectral estimation: a tunable high-resolution
  spectral estimator,''
\newblock {\em IEEE Transactions on Signal Processing}, vol. 48, no. 11, pp.
  3189--3205, 2000.

\bibitem{Xu2008}
L.~Xu, J.~Li, and P.~Stoica,
\newblock ``Target detection and parameter estimation for {MIMO} radar
  systems,''
\newblock {\em IEEE Transactions on Aerospace and Electronic Systems}, vol. 44,
  no. 3, pp. 927--939, 2008.

\bibitem{DelphiESR}
``{Delphi Automotive Radar: ESR 2.5},''
  http://www.autonomoustuff.com/delphi-esr-25.html,
\newblock Accessed: 2015-5-23.

\bibitem{Nelder1965}
J.~Nelder and R.~Mead,
\newblock ``A simplex method for function minimization,''
\newblock {\em Computer Journal}, vol. 7, pp. 308--313, 1965.

\bibitem{Riihonen2012}
T.~Riihonen and R.~Wichman,
\newblock ``Analog and digital self-interference cancellation in full-duplex
  {MIMO-OFDM} transceivers with limited resolution in {A/D} conversion,''
\newblock in {\em Proceedings of the Conference Record of the Forty Sixth
  Asilomar Conference on Signals, Systems and Computers}, 2012, pp. 45--49.

\bibitem{Bharadia2013}
D.~Bharadia, E.~McMilin, and S.~Katti,
\newblock ``Full duplex radios,''
\newblock in {\em ACM SIGCOMM Computer Communication Review}, 2013, pp.
  375--386.

\bibitem{Waveforms80211ac}
``802.11ac waveform generation with mac frames,''
  \url{https://www.mathworks.com/help/wlan/examples/802-11ac-waveform-generation-with-mac-frames.html},
\newblock Accessed: 2017-05-04.

\bibitem{Rappaport2002}
T.~Rappaport,
\newblock {\em Wireless communications: principles and practice},
\newblock Prentice-Hall, 2002.

\bibitem{Suzuki2000}
H.~Suzuki,
\newblock ``Measurement results of radar cross section of automobiles for
  millimeter wave band,''
\newblock in {\em Proceedings of the {7}th World Congress on Intelligent
  Systems}, 2000, p.~6.

\end{thebibliography}

\end{document}